\documentclass{vgtc}                          

\ifpdf
  \pdfoutput=1\relax                   
  \usepackage{graphicx}                
  \DeclareGraphicsExtensions{.pdf,.png,.jpg,.jpeg} 
\else
  \usepackage{graphicx}                
  \DeclareGraphicsExtensions{.eps}     
\fi%

\graphicspath{{figures/}{pictures/}{images/}{./}} 

\usepackage{microtype}                 
\PassOptionsToPackage{warn}{textcomp}  
\usepackage{textcomp}                  
\usepackage{mathptmx}                  
\usepackage{times}                     
\usepackage{cite}                      
\usepackage{tabu}                      
\usepackage{booktabs}                  

\usepackage{mathptmx}
\usepackage{graphicx}
\usepackage{times}
\usepackage{color}

\usepackage{amsmath}
\usepackage{amsfonts}
\usepackage{multirow}

\onlineid{0}

\vgtccategory{Research}

\vgtcinsertpkg

\title{Exploration of Heterogeneous Data Using Robust Similarity}

\author{Mahsa Mirzargar\thanks{e-mail: mirzargar@cs.miami.edu}\\ %
        \scriptsize University of Miami %
\and Ross T. Whitaker\thanks{e-mail:whitaker@cs.utah.edu}\\ %
     \scriptsize University of Utah %
\and Robert M. Kirby\thanks{e-mail:kirby@cs.utah.edu}\\ %
	\scriptsize University of Utah %
     }

\abstract{Heterogeneous data pose serious challenges to data analysis tasks, including exploration and visualization. 
Current techniques often utilize dimensionality reductions, aggregation, or conversion to numerical values to analyze heterogeneous data. However, the effectiveness of such techniques to find subtle structures such as the presence of multiple modes or detection of outliers is hindered by the challenge to find the proper subspaces or prior knowledge to reveal the structures. 
In this paper, we propose a generic \emph{similarity-based} exploration technique that is applicable to a wide variety of datatypes and their combinations, including heterogeneous ensembles. The proposed concept of similarity has a close connection to statistical analysis and can be deployed for summarization, revealing fine structures such as the presence of multiple modes, and detection of anomalies or outliers. We then propose a visual encoding framework that
enables the exploration of a heterogeneous dataset in different levels of detail and provides insightful information about both global and local structures. We demonstrate the utility of the proposed technique using various real datasets, including ensemble data.
} 

\CCScatlist{ 
  \CCScat{G.3}{Probability and statistics}%
{Nonparametric statistics }{};
  \CCScat{I.3.8}{Computer Graphics}{Applications}{}
}

\begin{document}

\firstsection{Introduction}

\maketitle


Data analysis is an indispensable component of knowledge acquisition. Oftentimes, scientists, engineers, and decision-makers employ exploratory data analysis to find any potential patterns or structures in the data intuitively before delving into more concise analysis.  One of the important aspects of today's data exploration and analysis is the complexity and heterogeneity. Heterogeneous data are generated and analyzed in fields as diverse as simulation sciences, medical applications, or meteorology. Effective analysis of heterogeneous data  requires proper and unified treatment of diverse and often incompatible datatypes present in a dataset. However, such generic data analysis techniques are mainly lacking. Therefore, visualization can provide a valuable tool to facilitate data exploration for new discoveries and the formation of new hypotheses. If designed well, visualization can help domain experts to explore heterogeneous datasets more effectively.

The goal of this work is to introduce a simple and generic concept of similarity that has a close connection with concepts from descriptive statistics. The proposed definition is applicable to a variety of datatypes and their combinations and can be deployed to 
analyze and explore a heterogeneous dataset at various levels of abstractions, namely: global trends, presence of potentials modes, and detection of outliers. 
Following Shneiderman's information-seeking mantra, we present a visual encoding technique that enables visualization of a heterogeneous dataset in a hierarchical fashion to reveal both global and local trends robustly. For datasets that do not have any geospatial attribute, our proposed visual encoding technique includes a self-organizing layout of the dataset combined with a set of interactive tools supporting user-driven exploration. 

The contributions of this work are as follows:
\begin{itemize}
\item A generic pairwise measure of similarity based on the concept of containment applicable to mixed-typed or heterogeneous data.
\item A robust method for detecting the presence of fine structures such as the presence of multiple modes or outliers using the proposed concept of similarity and spectral clustering. 
\item A visual encoding technique that provides multiple linked views, including an overview layout, a heatmap visualization of the similarity matrix between various data points.
\end{itemize}


\section{Related Work}
\label{sec:bg1}

The visualization community has been working on the problem of heterogeneous data visualization from different aspects. In what follows, we provide a brief introduction to some of the work relevant to our contribution. 

One aspect of heterogeneity that has been well studied before is the challenges related to analyzing and visualizing data from multiple sources~\cite{Uselton98}. Various visualization techniques have targeted such heterogeneity. For example, StratomeX~\cite{Lex12}, a visual analysis tool for heterogeneous genomics data visualization, is particularly designed for comparative visualization of heterogeneous datasets that share at least one common identifier. Pathline~\cite{Meyer10} is another example that facilitates comparative analysis of quantitative values from different data sources. Kehrer et al. targeted the problem of heterogeneity in scientific data and designed a visualization system that has enabled the users to visually explore data consisting of two or more data parts through the use of carefully designed coordinated multiple views~\cite{kehrer2011interactive}. Hybrid reality has also been utilized to visualize large heterogeneous datasets~\cite{reda2013visualizing}. Cammarano et al. proposed a visual analytics system in which they focus on querying and visualizing loosely coupled heterogeneous data. To represent data objects in their system, they perform a feature matching to select attributes that match a given visualization specification~\cite{cammarano2007visualization}. Unlike this class of techniques that mainly focus on multisource aspect of a heterogeneous data, and how one can maintain the integrity of the data while looking at data from various sources, our work focuses on mixed-datatype aspects of a heterogeneous data.

Other techniques that are closely related to our work are the group of visual analysis techniques that consider a heterogeneous or mixed-type dataset as high-dimensional data. The information and scientific visualization literature provides a variety of techniques to visualize high-dimensional data~\cite{Liu15}. Many of the high-dimensional visualization methodologies that have been utilized for heterogeneous data visualization can be viewed as dimensionality reduction techniques. 
Start coordinates~\cite{Kandogan00}, RadViz~\cite{Sanchez16} and techniques based on multidimensional scaling~\cite{Williams04} are among the most popular techniques that benefit from either linear or nonlinear dimensionality reduction techniques. 
Studies have shown that Star coordinates can be a powerful technique for exploring specific structures in high-dimensional data such as the presence of clusters or outliers~\cite{Sanchez16}. However, this capability requires finding axis vectors where data reveal its underlying structures, which can be challenging to do automatically. 
Variations of Star coordinates such as RadViz take advantage of nonlinear mappings of high-dimensional data in order to find a proper subspace to represent the data. Nonlinear mapping is particularly useful when dealing with sparse datasets or datasets that cannot be represented effectively using linear transformations. However, the distortions introduced as part of the nonlinear mapping of data hinder the ability to estimate the original data attributes~\cite{Sanchez16}. 

Another class of techniques uses special visual mapping techniques to visualize high-dimensional datasets so that the user can observe the patterns or structure in the data intuitively. Scatterplots and parallel coordinates~\cite{Inselberg09} are among the prominent techniques in this group. Scatterplots or SPLOM are useful for visual detection of the correlation between two variables or finding clustering of datapoints in a dataset for which pairwise similarity or distance measures are available. One of the main concerns about SPLOM visualization is its scalability in terms of the size of the dataset and also depicting the relation between more than two variables (or dimensions). In comparison to scatterplots, parallel coordinates can provide a good overview of various attributes of high-dimensional data~\cite{Johansson16}. 
However, standard 2D parallel coordinates allow the identification of relationships only between adjacent axes. Therefore, the ordering of the axis plays a major role when the goal is to find structures in high-dimensional datasets~\cite{Johansson16}. 

Multidimensional scaling (MDS) is another popular technique for dimensionality reduction~\cite{Williams04} that tries to find the best low-dimensional projection of high-dimensional datapoints while preserving the similarity or distance between the datapoints in the original space. Similar to other dimensionality reduction techniques, MDS also loses individual dimensional information and can be very expensive. Moreover, MDS facilitates the exploration of data in a subspace that maximally preserves the original similarity or distance information, but it does not automatically detect any structures in the data;  detection has to be performed as a postprocessing stage. MDS has been utilized for heterogeneous data visualization through definition of a new distance metric called the structural similarity index~\cite{lee2014structure}. This index is a perceptual distance metric that has been utilized in conjunction with traditional Euclidean distance to enhance the MDS representation of the outcome of already-classified data. In addition to the notion of \emph{distance}, the notion of \emph{similarity} has also been used before for visualization of heterogeneous data. A data context map~\cite{cheng2016data} is an example of techniques that focus on finding similarity of elements in a dataset from various aspects, namely: the similarity of data objects, the similarity of attributes, and the relationship between data object and attributes. It then deploys an MDS-type analysis to find a lower dimensional representation in terms of a single comprehensive map. In this approach, any categorical attributes will be converted into numerical ones before analysis.

In this paper, we provide a more direct and generic approach for the analysis of heterogeneous (or mixed-type) datasets by defining a measure of pairwise similarity that is generic enough to be applicable to various datatypes and their mixtures. The proposed similarity concept relies on the simple concept of containment and, hence, does not require any datatype conversion. In addition, the proposed notion of similarity can robustly reveal the presence of multiple modes in a dataset if any are present. In the next section, we introduce the proposed notion of similarity in more detail. We then introduce a multiple-view visual encoding framework that enables the analysis and exploration of heterogeneous datasets in a hierarchical fashion while facilitating detection of both \emph{global features} such as trends and \emph{fine structures} such as the presence of multiple modes or outliers. 

%


\section{Technical Background}
\label{sec:bg2}

The main premise of our work relies on the \emph{pattern} of the inclusion of a datapoint \emph{between} a random subset of other datapoints in a dataset that we call \emph{band inclusion}. The concept of inclusion between random subsets relates to a known concept from descriptive statistics called \emph{band depth} to measure centrality. 
Since data depth is not the main focus of the current manuscript, we do not cover its definition and properties~\cite{Zuo00}. 

\subsection{Band Inclusion Signature}

For simplicity of the discussion and without loss of generality, we explain the basic idea in $1$D. 
Consider a set of univariate datapoints: $\mathcal{X} = \{ m_1, \cdots, m_n\}$. Let us define a \emph{band} $lB_{jk}$ to be the line segment between two data points from this set, $m_j$ and $m_k$. For any $m_i \in \mathcal{X} $, a band inclusion value can then be defined as 
\begin{equation}
B_{jk}(m_i) = I(m_i \subset lB_{jk}) 
\label{eq:inclusion}
\end{equation}
where $I(\cdot)$ denotes an indicator function and the right hand is defined as follows
\vspace{-5pt}
\begin{equation}
I(m_i \subset lB_{jk}) = \begin{cases}
1 & m_i \in [\min(m_k, m_j), \max(m_k, m_j)]\\
0 & \text{otherwise.}
\end{cases}
\label{eq:inclusion}
\end{equation}
Considering random bands constructed using random subsets, 
one can assign a \emph{band inclusion signature} to each datapoint, let us call it $s_i$. 
Hence, the band inclusion signature is a binary string whose size is the same 
as the number of random bands considered. For instance, if all the subsets of size 
two are considered for the band inclusion signature of our 1D example, each 
datapoint would be assigned a band inclusion signature of length $C(n,2)$ that can be represented as
\begin{equation}
s_i(j+nk) = I(m_i \subset lB_{jk}), \quad 1 \leq i,j,k \leq n, \; j > k,
\end{equation}
where $s_i(m)$ denotes the $m^{th}$ element 
of the band inclusion signature for the $i^{th}$ datapoint. 
\begin{figure}[h!]
  \begin{center}
  \begin{tabular}{@{\hspace{0pt}}c@{\hspace{0pt}}}
    \includegraphics[width=0.3\textwidth]{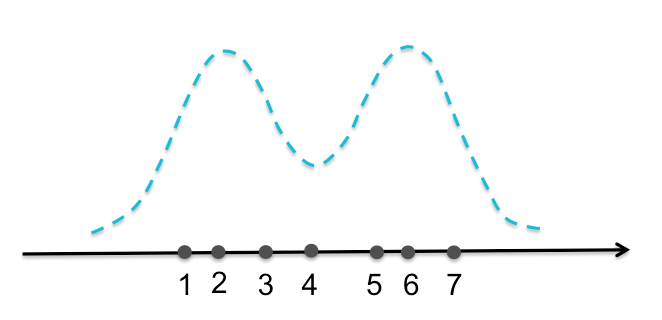}  \vspace{-3pt}\\
    (a) \\ 
     \includegraphics[width=0.4\textwidth]{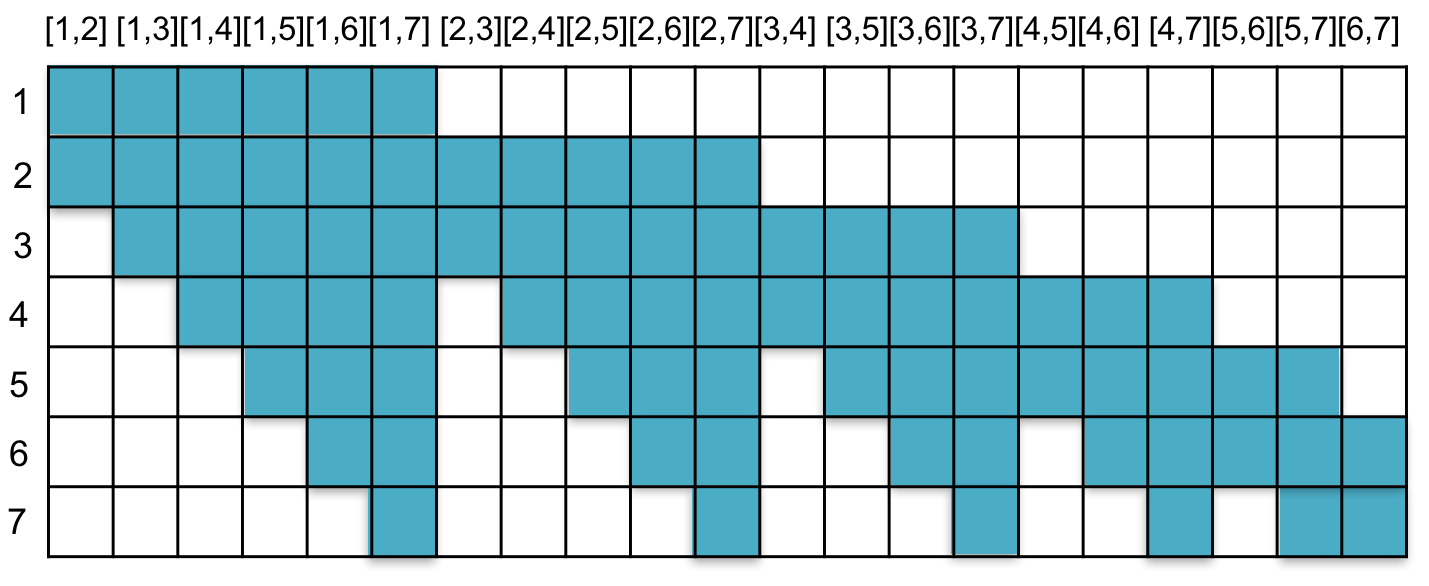}  \vspace{-3pt}\\
     (b) \\ 
    \end{tabular}
     \vspace{-5pt}
      \caption{\label{fig:sig} (a) A set of 1D points drawn from a bimodal distribution. (b) Each row represents the band inclusion signature using Eq.~\ref{eq:inclusion}. The numbers at the top of the columns correspond to the datapoint indices used to construct the band.}
      \end{center}
     \vspace{-10pt}      
\end{figure}
Figure~\ref{fig:sig} depicts the idea behind the band inclusion signature 
for a set of 1D points. The concept of band inclusion is data-type dependent and Table~\ref{ta:bd} provides the reference for the definitions of band inclusion for other popular datatypes~\cite{liu1990,Pintado06,Whitaker13,Mirzargar14}.

The band inclusion signature provides information 
regarding the centrality of each datapoint within a dataset. 
The more central a datapoint is, the more 
nonzero values it has in its band inclusion signature. A reader familiar with the concept of data 
depth will notice that the mean of the band inclusion signature 
is the data depth value for each datapoint, but the notion of band inclusion 
signature has never been used before to the best of our knowledge. 

In addition to information about centrality, the band inclusion signature
also provides information about the \emph{similarity} of different datapoints. 
The idea behind the measure of similarity based on the band inclusion signature is 
that datapoints more similar to each other are more likely to 
have a similar band inclusion signature. 
For instance, examples two and six have the same number of nonzeros in their 
band inclusion signature, but their band inclusion signatures look very different 
(the same is true for examples one and seven or three and six). 
In comparison to example two, examples five and six have more overlapping 
nonzero values in their band inclusion signature.

While the band inclusion signature provides both global information in terms of the centrality 
of each datapoint and the pairwise similarities between different datapoints, 
%
it suffers from the same drawback as data depth analysis. 
That is, it fails to provide reasonable information about the 
more subtle properties of a dataset such as the presence of multiple modes (if any). 
One way to alleviate this problem is to modify the band inclusion signature in such a way 
that it becomes sensitive to local features that might be present in a dataset. 
For instance, for a dataset with multiple modes or clusters, one would like 
to be able to define a similarity measure that would automatically reflect 
such information. In other words, the goal is to define a similarity measure 
that would consider datapoints in the same mode to be more similar compared 
to datapoints from different modes. We can achieve this goal by 
restricting  the \emph{size} of the random bands that are used 
for construction of the band inclusion signature and using only random bands with sizes 
no greater than a fixed threshold $\tau$~\footnote{Theoretically, the size of a band would be its volume measure.}. 
 The motivation for using a band size threshold is that the 
inclusion of datapoints inside \emph{large} bands will not be informative 
about local features such as the presence of multiple modes. 
The introduction of a tuning parameter for the size of the random 
bands to better capture local features was motivated 
by a similar technique that has been proposed to overcome the shortcomings 
of data depth analysis of datasets generated from a multimodal 
distribution called \emph{local depth}~\cite{Agostinelli11}. 
The interested reader can refer to~\cite{Agostinelli11} 
for a detailed and theoretical discussion~\footnote{The proof of convergence of local 
depth maximizers to the modes of a probability 
distribution (or an ensemble drawn from it) directly applies 
to the modified band inclusion signature.}.  

As an example, the definition of $s_i$ (Eq.~\ref{eq:inclusion}) can be 
modified as follows for our 1D example:
\begin{equation}
s_i(j+nk) = \begin{cases}
 I(m_i \subset lB_{jk}) & \text{ iff } |\min(m_k, m_j), \max(m_k, m_j)| \leq \tau \\
 0 & \text{ otherwise,}
 \end{cases} 
\label{eq:inclusion2}
\end{equation}
where $|\cdot|$ denotes the length of the interval $[\min(m_k, m_j), \max(m_k, m_j)] $. 
\begin{figure}[h!]
  \begin{center}
  \begin{tabular}{@{\hspace{0pt}}c@{\hspace{0pt}}}
     \includegraphics[width=0.4\textwidth]{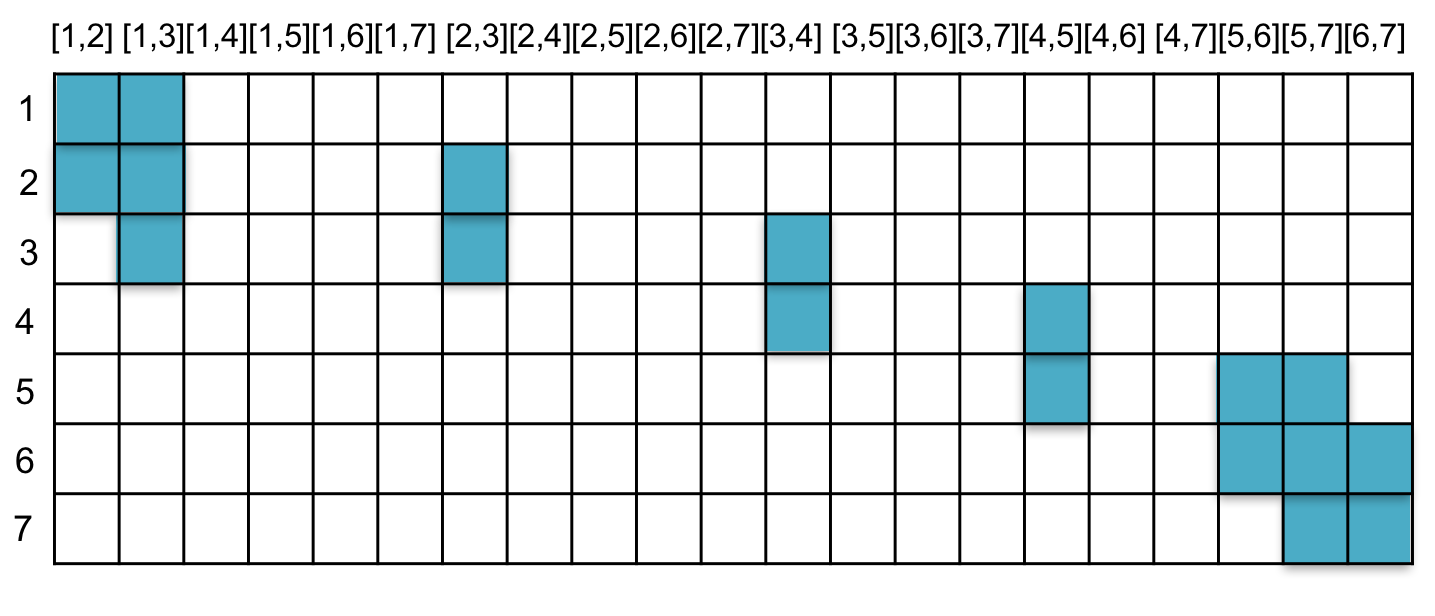}  \vspace{-3pt}\\
    \end{tabular}
     \vspace{-5pt}     
      \caption{\label{fig:sig2} Each row represents the band inclusion using Eq.~\ref{eq:inclusion2}. Similar to Figure~\ref{fig:sig}, the numbers at the top of the columns correspond to the datapoint indices used to construct the band.
 }
      \end{center}
           \vspace{-12pt}     
\end{figure}
Figure~\ref{fig:sig2} demonstrates the modified band inclusion signature where $\tau=1.3$ has been 
chosen. Note that unlike Figure~\ref{fig:sig}, examples two and six now have the highest 
number of nonzero values in their band inclusion signatures. 
\begin{table*}
   \begin{center}
   \scalebox{.85}{
   \begin{tabular}{|c|@{\hspace{5pt}}c@{\hspace{5pt}}|@{\hspace{1pt}}c@{\hspace{1pt}}|@{\hspace{1pt}}c@{\hspace{1pt}}|c|}
   \hline
   Datatype & Corresponding Depth Concept & Band Inclusion Definition & Band Size \\ \hline
   $n$-d Multivariate Points & Simplicial Depth & $x \subset B_{\Delta}(x_1, \cdots, x_{n+1}) \text{ iff} $ & \multirow{2}{*}{Vol$\{\Delta(x_1, \cdots, x_{n+1})\}$} \\
   ($x \in \mathbb{R}^n$) & ~\cite{liu1990} & $x \in \Delta(x_1, \cdots, x_{n+1})$ ($n$-d simplex) &   \\ \hline
   Functions & Functional Band Depth & $\displaystyle f \subset B_f(f_1, \cdots, f_j) \text{ iff } \forall x \in \mathcal{D},$ & Functional Integration of  \\
   ($f: \mathcal{D} \mapsto \mathcal{R}$) & ~\cite{Pintado06} & $\displaystyle  \min_{k=1,\cdots, j} f_k(x) \leq f(x) \leq \max_{k=1,\cdots,j} f_k(x)$ & $B(f_1, \cdots, f_j)$~\cite{Mirzargar16} \\ \hline
   Sets (or Categorical) & Set Band Depth & $S \subset sB(S_1, \cdots, S_j) \text{ iff}$ & Power Set of  \\
   ($S \subset U$, $U$ universal set) & ~\cite{Whitaker13} & $\displaystyle \bigcap_{k=1}^{j} S_k \subset S \subset \bigcup_{k=1}^j S_k$ & $sB(S_1, \cdots, S_j)$\\ \hline
   Curves or Streamlines & Simplicial Band Depth & $\displaystyle f \subset B_f^{\Delta}(f_1, \cdots, f_{d+1}) \text{ iff } \forall x \in \mathcal{D},$ & \multirow{2}{*}{$\displaystyle \prod_{\forall x \in \mathcal{D}}\text{Vol}\{\Delta(f_1(x), \cdots, f_{d+1}(x))\}$} \\
   ($f: \mathcal{D} \mapsto \mathcal{R}^d$) & ~\cite{Mirzargar14, pintado2014simplical}  & $\displaystyle   f_k(x) \in \Delta(f_1(x), \cdots, f_{d+1}(x))$ & \\
   \hline
   \end{tabular}}
   \caption{   \label{ta:bd} The definition of band inclusion and band size for various datatypes.}
   \end{center}
\end{table*} 
The definition of the band size depends on the datatype and correspondingly 
the definition of a band. Table~\ref{ta:bd} summarizes different definitions of band size for different datatypes. 

\begin{figure}
\begin{center}
\begin{tabular}{@{\hspace{0pt}}cc@{\hspace{0pt}}}
\includegraphics[width=0.15\textwidth]{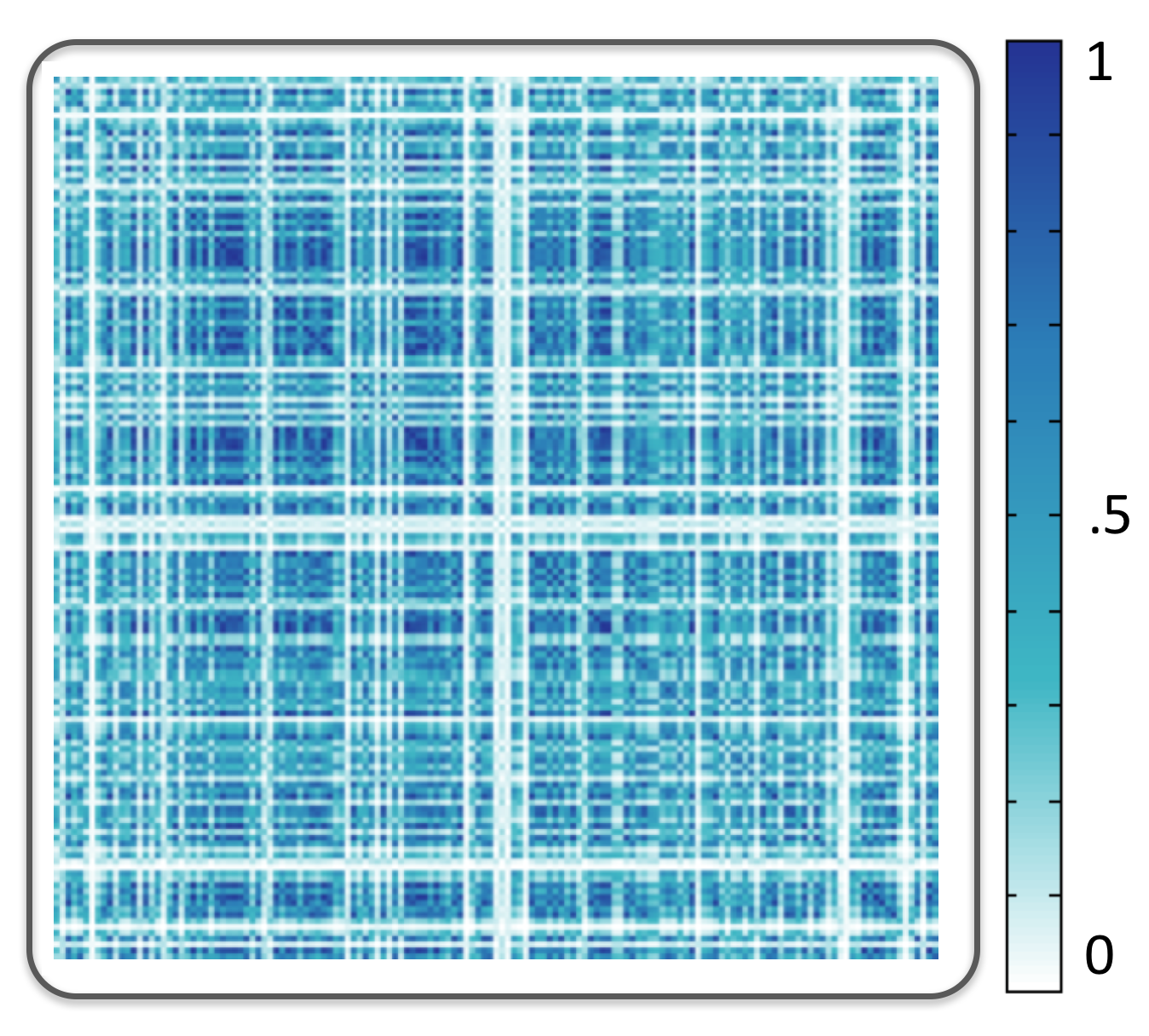} & \includegraphics[width=0.15\textwidth]{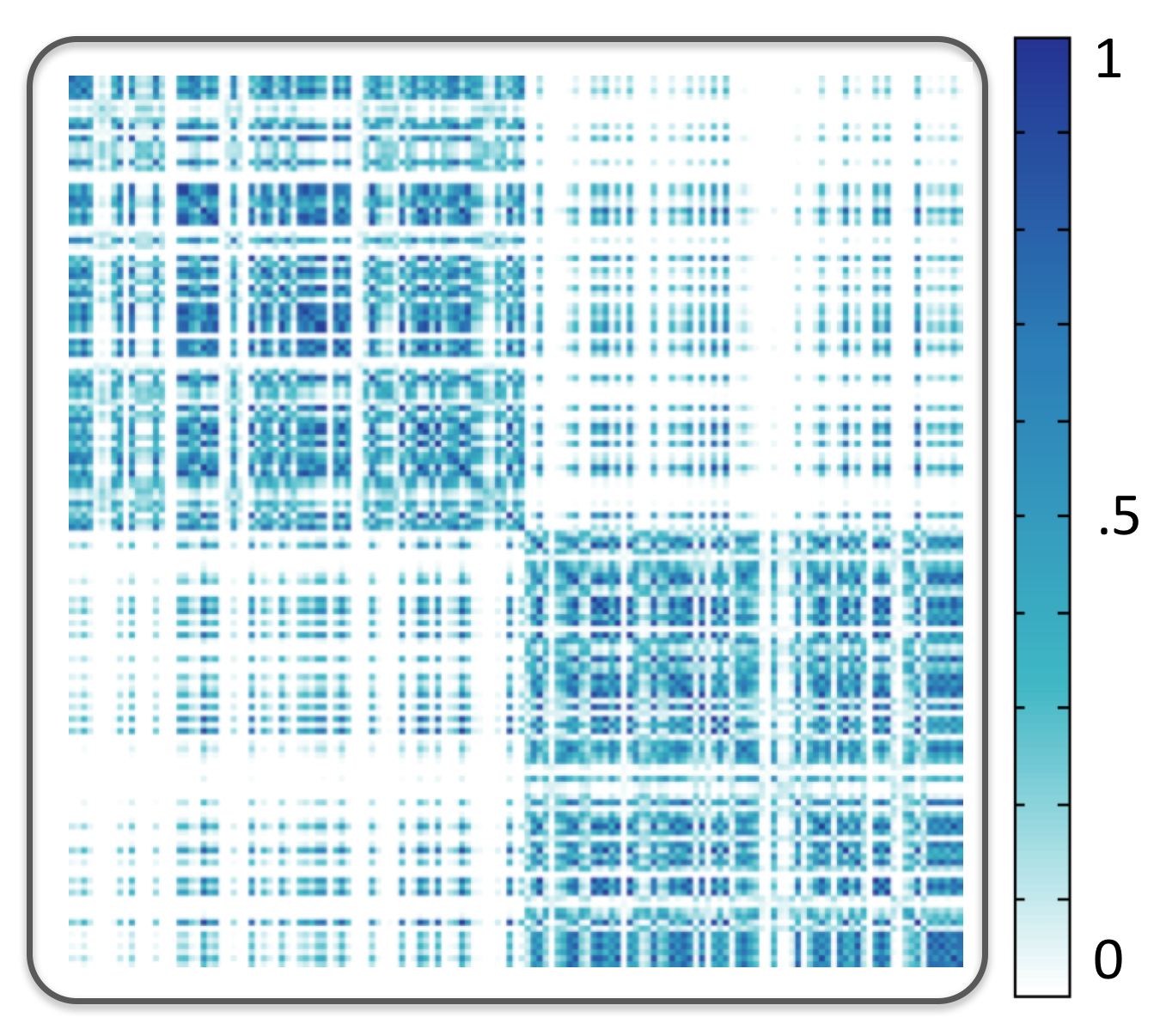}  \vspace{-3pt}\\
\includegraphics[width=0.15\textwidth]{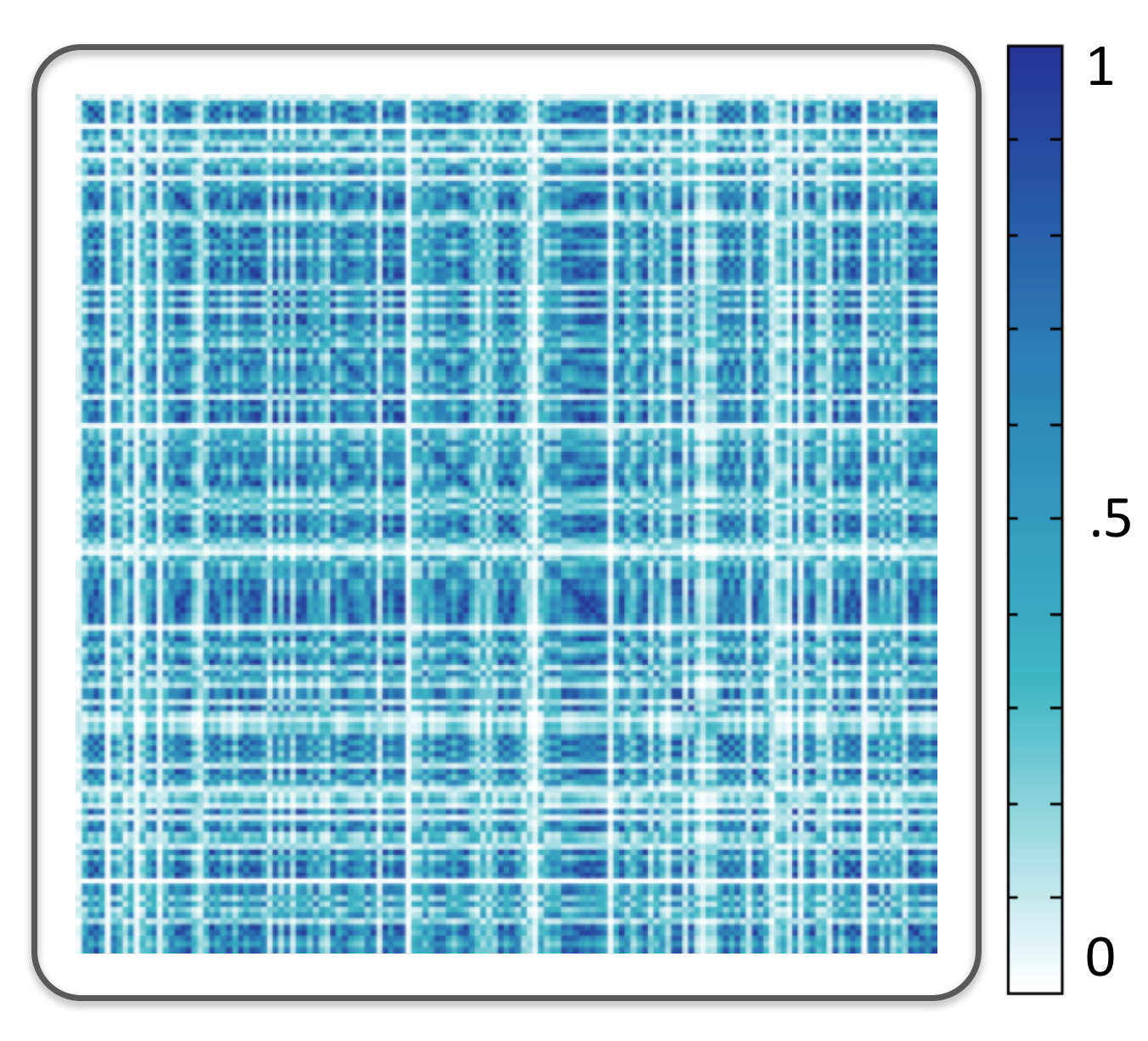} & \includegraphics[width=0.15\textwidth]{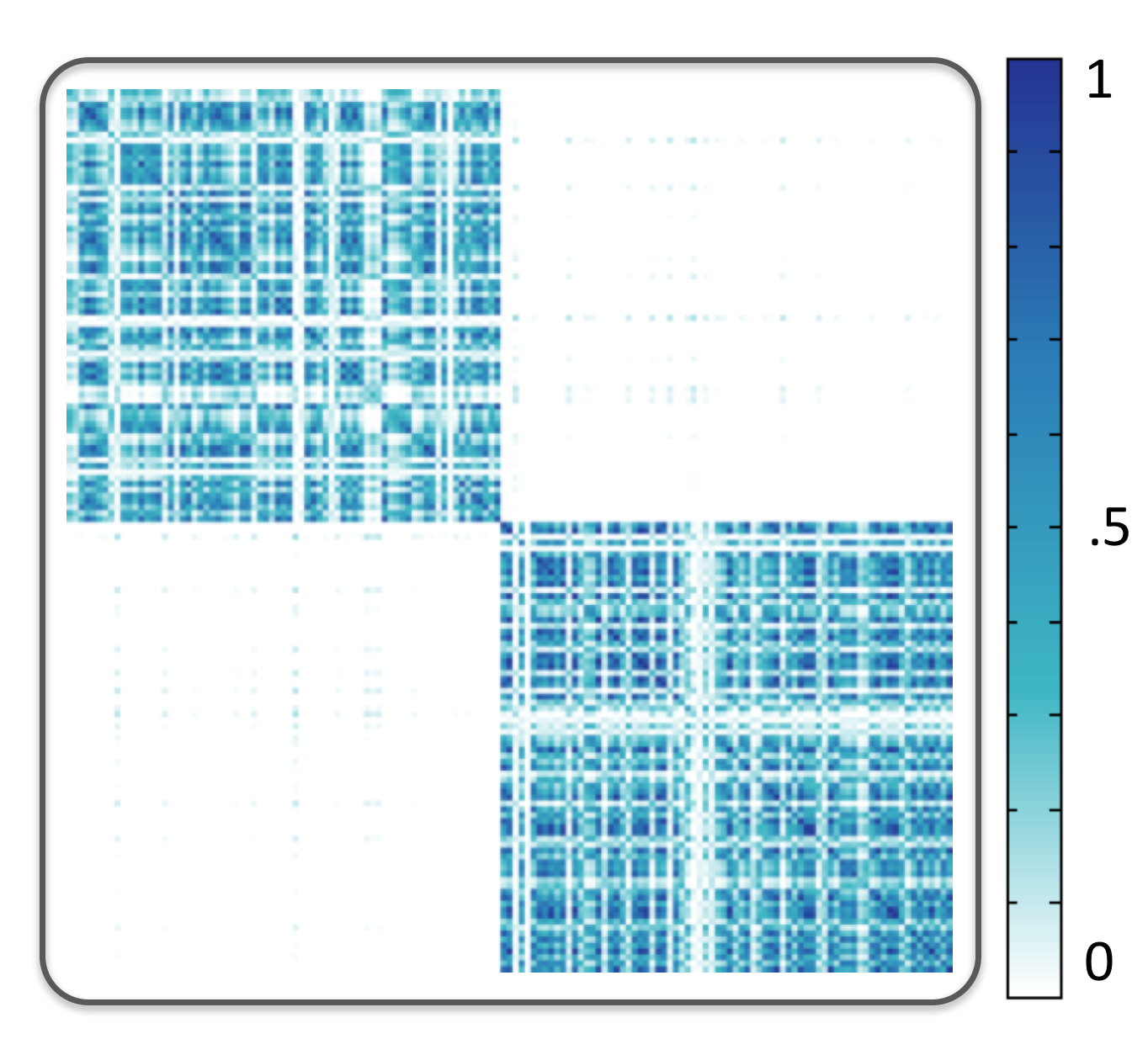}  \vspace{-3pt}\\
without band size thresholding & with band size thresholding \\
\end{tabular}
 \vspace{-5pt}
\caption{\label{fig:modal} Similarity matrix of an ensemble of size $99$ from a unimodal distribution (first row) and a bimodal distribution (second row). Both heatmaps on the right have been reordered based on the spectral clustering results.}
\end{center}
 \vspace{-12pt}
\end{figure}
It is important to note that the band inclusion signature has robust behavior with respect to the tuning parameter $\tau$. That is, if a dataset has been generated from a unimodal distribution, changing the value of $\tau$ \emph{will not} result in the introduction of multiple modes (see Figure~\ref{fig:modal}). This behavior is a direct result of the robustness of rank statistics induced using data depth values (i.e., the mean of the band inclusion signature). In other words, if one uses the measure of centrality, or the number of nonzero values in the band inclusion signature, to rank the datapoints, the ranking remains the same for datasets drawn from a unimodal distribution for different values of $\tau$. It is only for datasets with multiple modes that changing $\tau$ will reveal the multimodal natures. Discussion of the theoretical proofs 
is beyond the scope of the current manuscript, and we refer the interested reader to nominal works in the statistical literature for detailed discussions~\cite{Agostinelli11}.  Therefore, the parameter $\tau$ can be considered as a tuning parameter that progressively reveals the structures present in a dataset. This behavior is in contrast to widely used kernel-based techniques such as kernel principal component analysis or radial basis functions where varying the bandwidth of the kernel can affect the analysis result. This property also makes band inclusion analysis a more suitable technique for the exploration of datasets in different levels of detail compared to clustering algorithms where the number of clusters or modes needs to be specified a priori~\cite{Ferstl16}. 

In the presence of a heterogeneous dataset, the concept of band inclusion can easily be applied to each datatype separately. For instance, 
consider a dataset that includes datapoints consisting of two datatypes,  
univariate value ($x_i$) and a set membership ($s_i$): $\mathcal{X} := \{(x_1, s_1), \cdots, (x_n, s_n)\}$. For such a dataset, the band inclusion 
can be defined as
\begin{equation}
(x_i, s_i) \subset B_{jk} \leftrightarrow \{x_i \subset lB_{jk} \wedge s_i \subset sB_{jk}\},
\end{equation}
where $lB_{jk}$ is defined as in Eq.~\ref{eq:inclusion} and $sB_{jk}$ is defined using Table~\ref{ta:bd}. 
The band size in this case would be simply defined as the multiplication of 
the size of each constituent band as: $(|lB_{jk}| \cdot 2^{|sB_{jk}|})$ where 
$2^{|sB_{jk}|}$ denotes the size of the power set of $sB_{jk}$ (see Table~\ref{ta:bd}). Similar definitions 
can be derived for any other heterogeneous dataset. 

In the next section, we discuss how the inclusion signature and the similarity measure induced by it can 
be used to detect the presence of multiple modes or clusters automatically. 

\subsection{Constructing Similarity Matrix and Clustering}
\label{sec:clustering}

As discussed in the previous section, the \emph{pattern} of the nonzero values 
in the band inclusion signature can be used to measure the similarity between two datapoints in a dataset. 
More specifically, we define a measure of pairwise similarity 
as the distance between the band inclusion signatures. 
A band inclusion signature is nothing but a binary string; hence, we adopted hamming distance as a conventional distance measure for two binary strings to compute the distance between two band inclusion signatures. Hamming distance, a distance metric widely used in information theory, is defined as follows: the hamming distance between two strings of the same length is the number of positions at which the corresponding strings are different. 
By computing the hamming distance between each pair of band inclusion signatures in a dataset, we can construct a symmetric similarity matrix as 
\begin{equation}
\mathbf{S} = \bigg[1-s_{ij}\bigg]_{i,j=1}^n,
\label{eq:simmat}
\end{equation}
where $s_{ij}$ is the normalized hamming distance between the band inclusion signature of $i^{th}$ and $j^{th}$ datapoints. 

\begin{figure}
\begin{center}
\begin{tabular}{@{\hspace{0pt}}cc@{\hspace{0pt}}}
\includegraphics[width=0.15\textwidth]{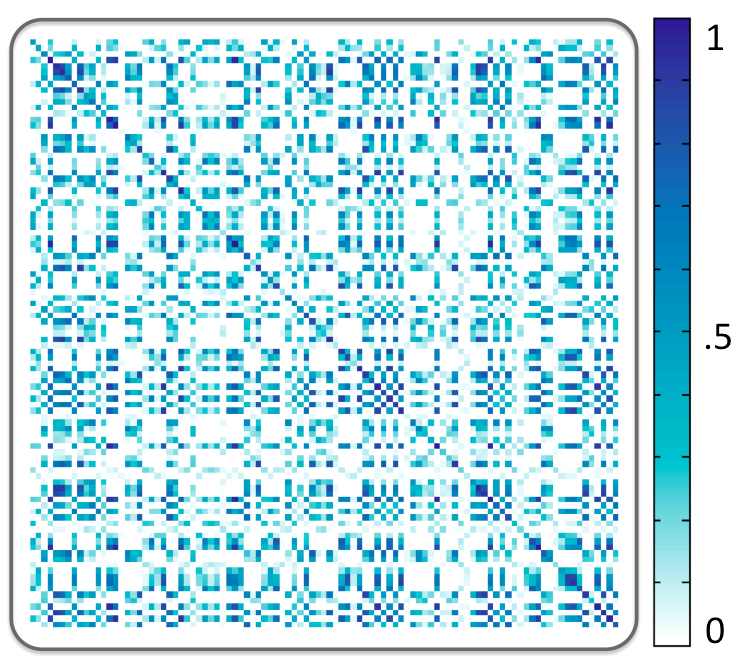} & \includegraphics[width=0.15\textwidth]{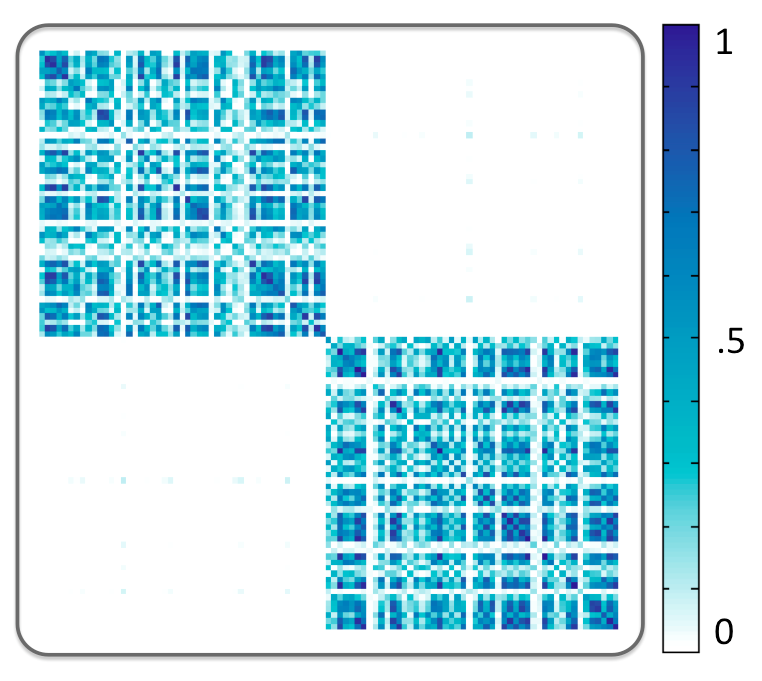}  \vspace{-3pt}\\
(a) & (b)
\end{tabular}
 \vspace{-5pt}
\caption{\label{fig:spectral} (a) Similarity matrix of an ensemble of size $99$ from the distribution shown in Figure~\ref{fig:sig} (a) using Eq.~\ref{eq:inclusion2}. (b) Reordering of the similarity matrix after performing spectral clustering clearly reveals the multimodal nature of the ensemble.}
\end{center}
 \vspace{-10pt}
\end{figure}
The similarity matrix is a powerful abstraction that 
can reveal any potential clustering when the rows (columns) are ordered properly. This idea has been well studied in the machine learning and visualization community and is the motivation for a class of clustering technique called \emph{spectral clustering}~\cite{Berkhin06,schultz2013open}.

We use spectral clustering for ordering the rows (columns) of the proposed similarity matrix in order to reveal any potential modality information in a dataset  (see Figure~\ref{fig:spectral}). 
In what follows, we focus on providing a short introduction to the spectral clustering methodology in order to make this section self-contained.

Given a set of datapoints $\{x_1, \cdots, x_n\}$ and information about the pairwise similarity, one can represent the relationship between the datapoints in terms of a weighted graph $\mathbf{G}(X,E)$ where each node $x_i$ in the graph represents the corresponding datapoint. 
Two nodes $x_i$ and $x_j$ on this graph are connected only if the similarity $s_{ij}$ between the corresponding datapoints is nonzero and the corresponding edge is weighted based on the similarity value. For simplicity, we can represent the graph $\mathbf{G}$ in terms of its adjacency matrix $\mathbf{S} = [s_{ij}]$; that is, the similarity matrix introduced in Eq.~\ref{eq:simmat}. 
The clustering problem can now be cast as finding the partitioning or \emph{cut} of this graph in such a way that the edges connecting various partitions have a minimum weight (and correspondingly the edges within each partition have high weights). 
The minimum cut of this graph identifies the optimal partitioning of the dataset. Finding the minimum cut is an NP-hard problem. Hence, graph Laplacian~\cite{Malik00} is used to find an approximate solution efficiently. In short, the graph Laplacian of $\mathbf{G}$ is defined as
\begin{equation}
\mathbf{L} = \mathbf{I} - \mathbf{D}^{\frac{-1}{2}} \mathbf{S} \mathbf{D}^{\frac{-1}{2}}.
\end{equation}
$\mathbf{D}$ is the degree matrix of the graph where each diagonal element is defined as $d_i = \sum_{j=1}^n s_{ij}$ and $\mathbf{I}$ represents the identity matrix.

The basic idea behind spectral clustering is to use the spectrum of the graph Laplacian (i.e., the eigenvalues and their corresponding eigenvectors) to define a proper partitioning of the graph and consequently the dataset into clusters (if any are present). The footprint of the eigenvalues, and correspondingly the eigenvectors of the Laplacian matrix, can provide insight into whether the dataset lends itself to clustering~\cite{Malik00}. This is one of the main motivations for using spectral clustering in this application since in many real-life applications, it is not obvious whether the dataset is generated from a multimodal distribution. 
In its simplest form, spectral clustering uses the second smallest eigenvector of the normalized graph Laplacian to partition the graph into two regions. In order to define $k$ clusters, one can use a $k$-means clustering of the eigenvectors. It is worth noting that 
in all the applications discussed in the subsequent section, we use the normalized Laplacian, which emphasizes the intercluster similarity~\cite{Malik00}.

In the next section, we introduce a visual encoding framework to visualize various features of a heterogeneous dataset learned from the band inclusion signature and spectral clustering.

\section{The Proposed Visual Encoding Technique}
\label{sec:method}

In this section, we propose a visual encoding technique that follows Shneiderman's information seeking mantra~\cite{shneiderman1996eyes}, that is, \emph{overview first, zoom and filter, details-on-demand}. One of the most effective visualization paradigms that has proven capabilities in data exploration is multiple coordinate views~\cite{Baldonado00}. Studies have shown that a combination of visual encodings has advantages in revealing the underlying structures, especially in the case of high-dimensional data. For instance, Holten et al.~\cite{Holten10} showed the efficacy of the combination of parallel coordinates and scatterplots for identification of clusters. Yuan et al.~\cite{Yuan09} proposed the concept of scattering points in parallel coordinates as a seamless integration of both techniques. Andrienko et al.~\cite{Andrienko10} also used a combination of self-organizing maps (SOM) and cartographic map display for the exploration of spatiotemporal patterns. 

At a high level, our proposed visual encoding technique consists of multiple linked views that we implemented in a prototype system (see Figure~\ref{fig:mushroom} and Figure~\ref{fig:sandy}).  
It is important to note that even though the individual components of our visual encoding framework rely on the existing high-dimensional data visualization techniques, the novelty of the proposed framework is to use the strength of each of these existing techniques combined with pairwise similarity analysis to facilitate the exploration of heterogeneous data to reveal insights about the underlying structure of a dataset at different levels. 
Our visual encoding framework includes: 1) a self-organizing layout of the dataset, if no geospatial information is present in the dataset, 2) a heatmap visualization of the similarity matrix, and 3) a panel for visualization of the attributes of the datasets at the bottom of the window. The visualization scheme used for each attribute depends on its datatype. 
We also provide a histogram visualization of the band sizes discussed in Section~\ref{sec:bg2} and an interactive slider to control the tuning parameter $\tau$. Interactive changes in one of the views during the data exploration will be reflected in other views, as discussed in more detail below. This framework allows users to maintain context across different levels of detail and explore the dataset and its properties in multiple ways. 

\subsection{Pairwise Similarity View}
The pairwise similarity between datapoints provides a high-level abstraction of the dataset. In this view, we provide a heatmap visualization of the pairwise similarities. The color in this view encodes the amount of similarity between pairs of datapoints in the dataset. The darker the color, the higher the similarity between pairs of datapoints. Revealing any structures in a dataset using a heatmap visualization very much depends on the ordering of the rows and columns (see Figure~\ref{fig:spectral}). Therefore, we use ordering based on the spectral clustering discussed in Section~\ref{sec:clustering} to order the rows and columns. 

\subsection{Similarity-Induced Layout View}

This view tries to provide a holistic perspective of the dataset and the goal is to: 1) provide a good overview of the full dataset, 2) signify any underlying structure of the data with minimal data manipulation (e.g., reordering required in the heatmap view), and 3) demonstrate the many-to-many relations between various datapoints. Since we are dealing with a heterogeneous dataset and we do not restrict the datatypes, this view needs to be flexible enough to handle common datatypes. Therefore, we consider two cases here. 

First, if the dataset includes any positional information (e.g., one of the attributes is location, isocontours, curves, etc.), we use the underlying (geospatial) embedding and the color channel in order to reveal the structures present in the dataset induced by the band inclusion similarity measure. The color channel encodes the number of nonzero values in the band inclusion signature of each datapoint. 
As discussed in Section~\ref{sec:bg2}, the number of nonzero values in the band inclusion signature can be considered as a measure of the centrality or representativeness of each datapoint. 
The darker the color, the more nonzero values the datapoint has in its signature. The reader familiar with the concept of data depth and boxplot visualization~\cite{Whitaker13} can consider the coloring scheme to be a depth coloring. Depending on the value for band size, $\tau$, the median (i.e., the global most central datapoint) or center of potential modes (i.e., the center of the local features) will be highlighted. 

Second, if the dataset does not include any positional information, in order to achieve the goals mentioned above, we need to design a layout that provides insight about the underlying structure. We also need to demonstrate the pairwise and many-to-many relationship between various datapoints. Therefore, we have decided to represent each datapoint using a 2D point and adopted a force-directed layout~\cite{Perer08}. A force-directed layout has been chosen as an automatic layout of node-link diagrams that has the potential to reveal the multimodality information if any is present~\cite{Noack04}. We have chosen to use the model that employs vertex-vertex repulsion and considers a spring force or edge weight between every two nodes of the graph. The edge weights have been assigned in such a way that the more similar two nodes are, the higher the spring force between them. Therefore, the spatial layout of the nodes or datapoints reveals any potential multimodality information~\cite{Noack04}. It is worth noting that there is no guarantee of convergence of the force-directed layout to the optimal positioning of the nodes. However, the layout can help reveal the multimodality information~\cite{Noack04}, which would reconfirm any pattern showing up in the heatmap visualization of the similarity matrix. In order to reduce the amount of overlap between nodes, we use quad-tree-based collision avoidance~\cite{Ghoshray96} after computing the automatic layout to shuffle the nodes that are significantly overlapping each other. The coloring scheme in this case is similar to that in the first case.

In addition to revealing the presence of any potential clusters in the dataset, this view also demonstrates the presence of any potential outliers in the datasets. Any (potential) outlier will have limited or no similarity to the rest of the datapoints in the dataset and, therefore, it will be assigned a different color based on depth coloring~\cite{Whitaker13}. If a force-directed layout is used, it will be repelled from the rest of the nodes in this view. In the presence of outliers or datapoints that have zero similarity to other datapoints, those outliers are  rendered close to the boundary of the window in an arbitrary position. The edge-drawing slider bar can be used to detect those outliers (if any).

\subsection{Secondary Views and Interaction Tools}

In addition to the two main views, we also include a barchart visualization of the histogram of the band sizes. The value of the band size threshold $\tau$ can be changed using the slider, based on which the other views will be updated. 
Being able to choose the band size threshold will enhance the exploration of both global and local features of the dataset. 

A slider bar at the bottom of the similarity-induced view is also provided to let the user choose a threshold for edge drawing when a force-directed layout is used (i.e., to help with clutter in case there exists a strong similarity between the datapoints). It is important to note that this slider controls only the edge drawing. Therefore, some of the points that appear to be disconnected can actually have nonzero similarity to the rest of the dataset, which can be explored by moving the slider.

To provide the capability of querying individual instances or datapoints from the dataset, we provide a separate panel for visualization of the data attributes. The visualization of the attributes depends on the datatype of the attributes. Categorical attributes are modeled as sets (see Table~\ref{ta:bd}), and hence, we use the stacked bar charts to represent the distribution of the datapoints among different values. The coloring of the bars follows the same coloring scheme used in the similarity-induced layout. For numerical values, both stacked bar chart or boxplot visualization and its generalizations can be used~\cite{Whitaker13, Mirzargar14}. In our examples, we have demonstrated both. 
The visual query of the data attributes is implemented using linking and brushing. Brushing any node in the self-organizing layout view highlights its corresponding attributes. In addition, brushing any specific attribute value highlights all the data elements with that specific value. Highlighting enables users to perform queries on each data element or attribute to better understand any potential association of specific data attributes to the structures revealed by the self-organizing layout and/or the depth coloring. 
In the next section, these capabilities will be better demonstrated using various examples.

\section{Results and Applications}
\label{sec:results}

\begin{figure*}[!ht]
  \begin{center}
  \includegraphics[width=0.7\textwidth]{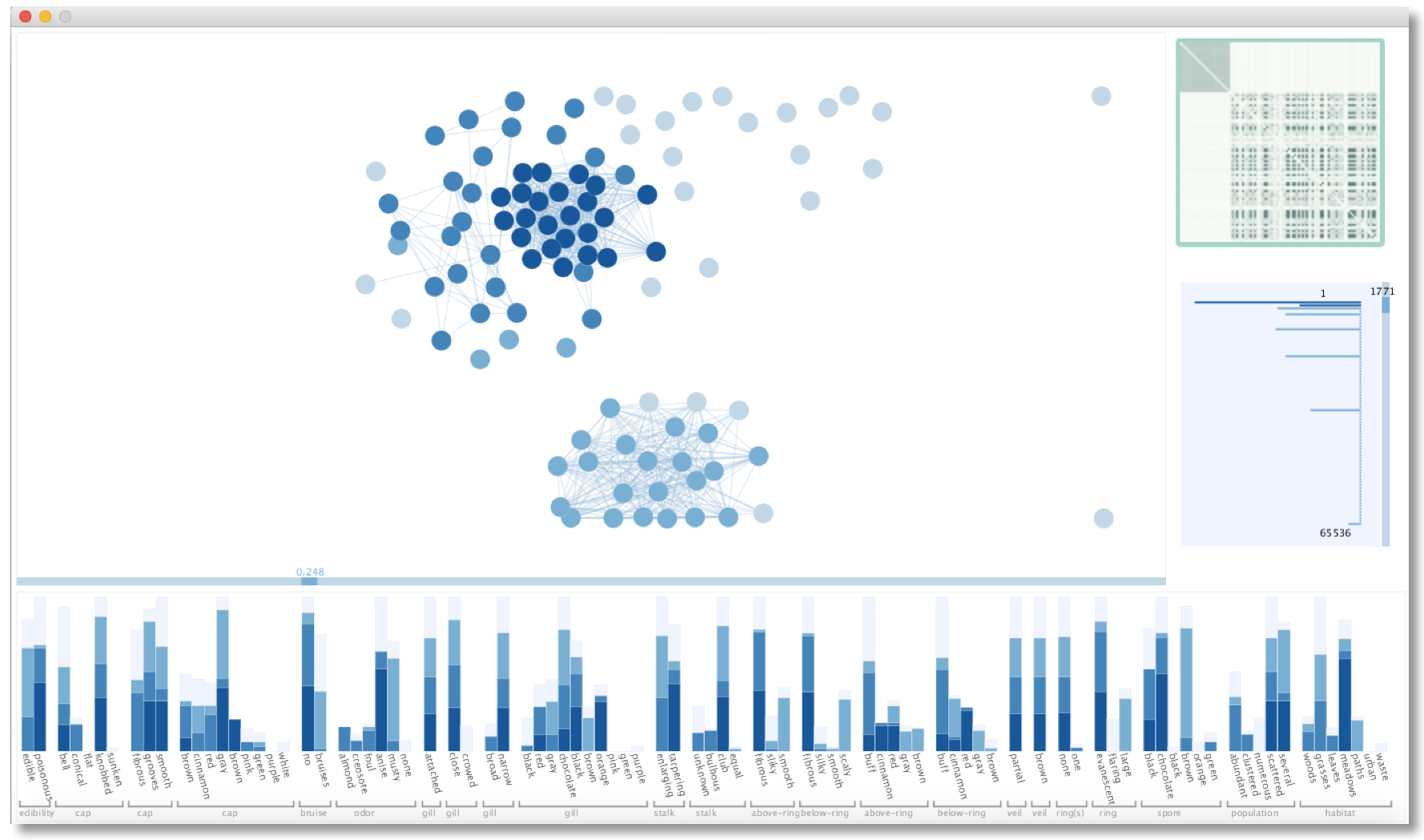}
  \vspace{-10pt}
      \caption{\label{fig:mushroom} Visualization of the mushroom dataset from UC Irvine Machine Learning Repository~\cite{mushroom}. The heatmap visualization of the pairwise similarities shows the presence of two clusters in this dataset, and the self-organizing layout provides more detail about the local information in the dataset and suggests the presence of three clusters.}
      \end{center}
  \vspace{-15pt}      
\end{figure*}
In this section, we demonstrate the utility of the proposed analysis and visualization technique using two examples. 
The band inclusion computation has been performed as a preprocessing step that was implemented in C++. The band inclusion computation for a dataset of size $250$ with $53$ attributes takes less than $30$ seconds on a single core machine. The band inclusion computation can be easily parallelized for large datasets. The results of the band inclusion computation (the signatures and band sizes) were then loaded into the visualization pipeline, which is implemented in Processing~\cite{processing}.

\begin{figure}[h!t]
  \begin{center}
  \begin{tabular}{@{\hspace{0pt}}c@{\hspace{0pt}}c@{\hspace{0pt}}}  
   \includegraphics[width=0.15\textwidth]{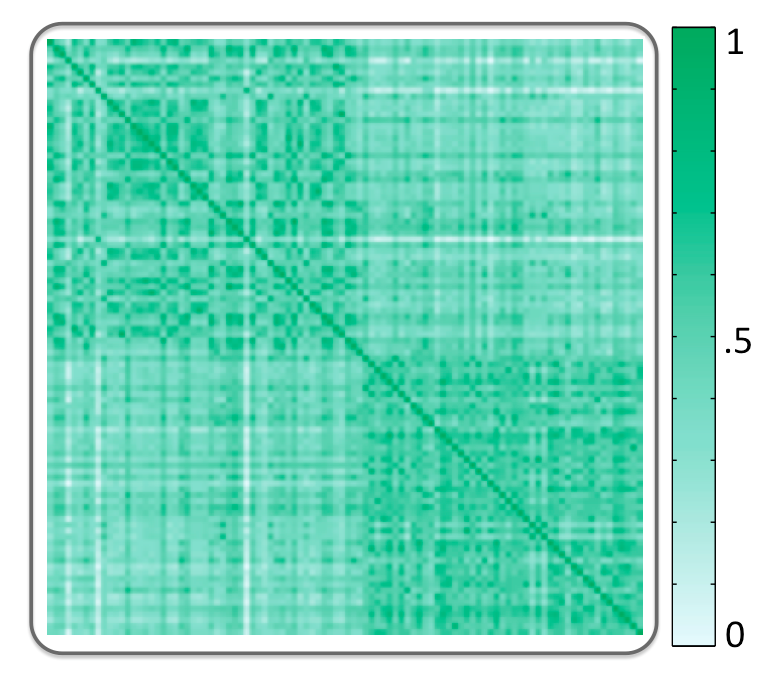} &     \includegraphics[width=0.15\textwidth]{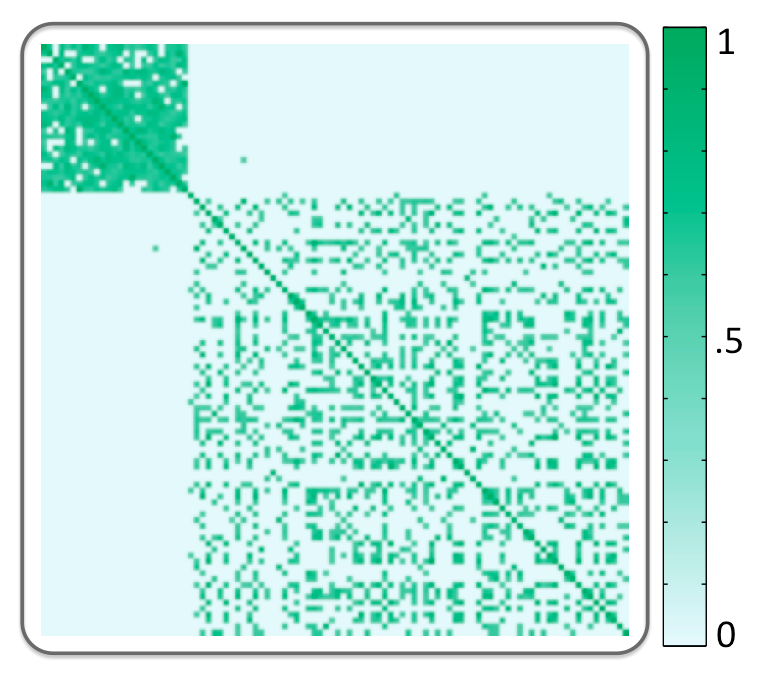}\vspace{-1pt}\\
 (a) & (c) \vspace{-2pt}\\
\includegraphics[width=0.24\textwidth]{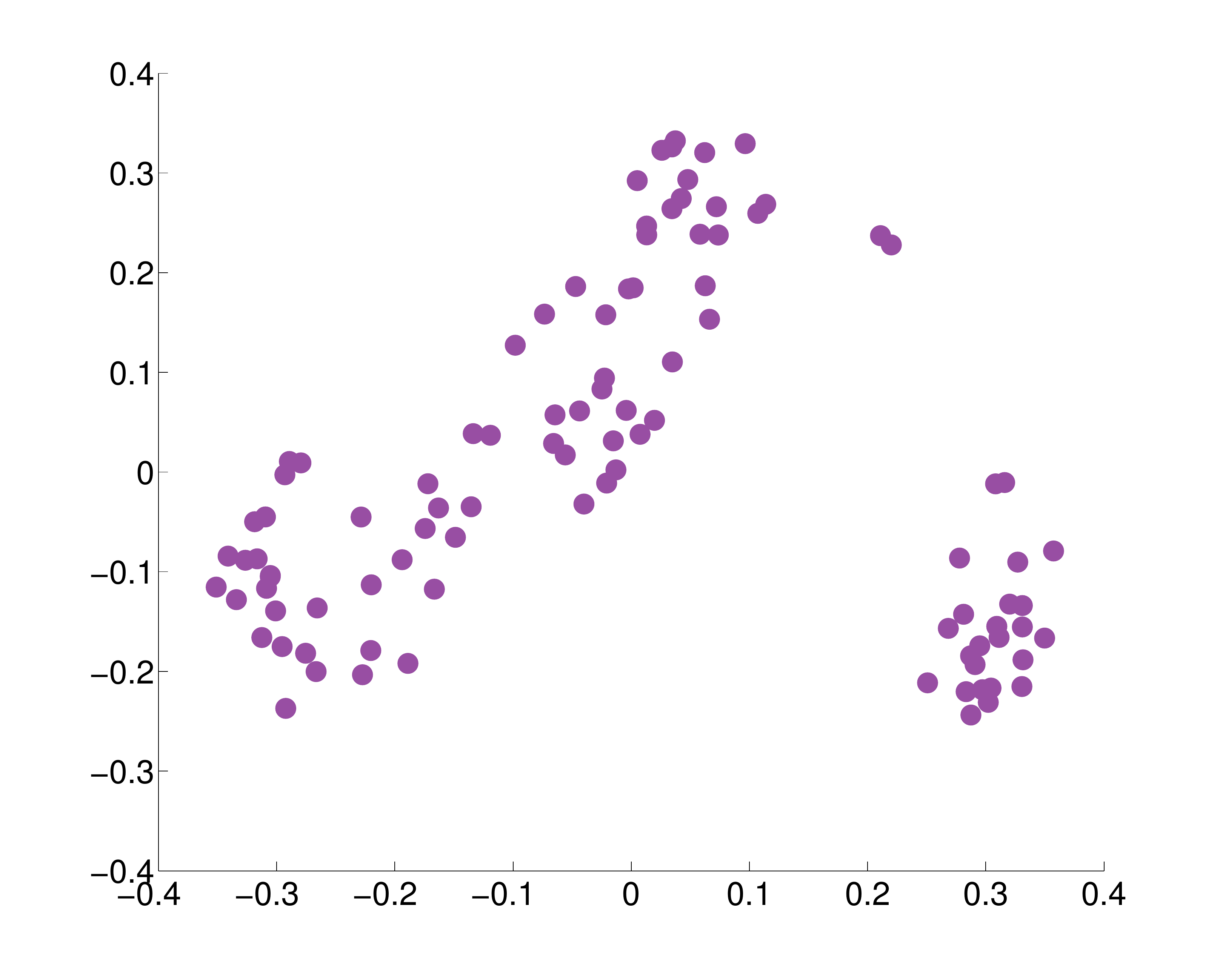} & \includegraphics[width=0.24\textwidth]{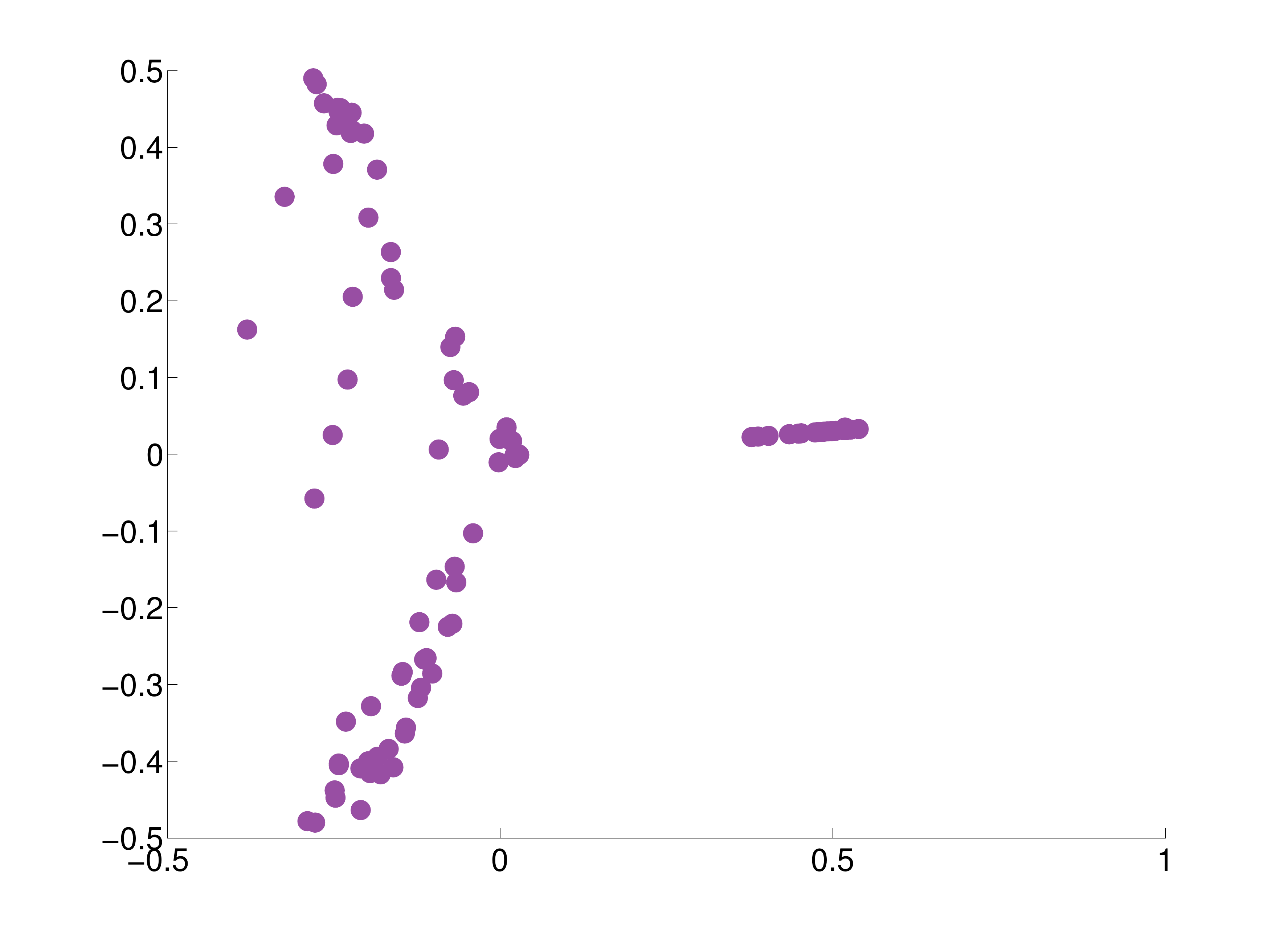} \vspace{-2pt}\\
(b) & (d) \\
    \end{tabular}
    \vspace{-5pt}
      \caption{\label{fig:mushroom_mds} Visualization of the mushroom dataset from UC Irvine Machine Learning Repository~\cite{mushroom} using Euclidean distance as the similarity measure and multidimensional scaling (MDS) as the dimensionality reduction technique. Thresholding the pairwise similarity values (c-d) will significantly affect both visualizations.}
      \end{center}
  \vspace{-10pt}      
\end{figure}
For our first example, we used the mushroom dataset from the UC Irvine Machine Learning Repository~\cite{mushroom}. This dataset includes $23$ attributes, with both categorical and numerical values. The categorical attributes mainly describe various physical properties of the mushroom: the cap shape, color, veil type, etc. The  numerical attributes determine the number of rings and whether the mushroom is poisonous or not. For more information about the attributes associated with this dataset, interested readers can consult~\cite{mushroom}. 
According to the description of the mushroom dataset, ``there is \emph{no} simple rule for determining the edibility of a mushroom''. Therefore, exploration of this dataset can better describe the properties of various types of mushrooms. 

Figure~\ref{fig:mushroom} demonstrates the visualization of the mushroom dataset for a dataset of size $100$. 
In this case, we have used four colors to represent $20\%$, $40\%$, $80\%$, and $100\%$ of the datapoints with the largest number of nonzero values in their signature. 
This color coding scheme tries to mimic a more detailed version of the color coding used in boxplot visualizations~\cite{Whitaker13, Mirzargar14}. 
By using the slider for the histogram of the band sizes, we can see that $\tau=1771$ (i.e, the band volume size) delineates two clusters in the heatmap view. For this specific value of $\tau$, the self-organizing layout suggests the presence of more than two subclusters. By hovering the cursor over the edibility attribute, it becomes clear that although edibility does not delineate these two clusters by itself, 
there exists a small cluster for edible mushrooms and a small cluster for which all the mushrooms are poisonous, whereas the bigger cluster of nodes is a mixture of edible and poisonous mushrooms. An interesting question to ask is whether any of the other attributes can delineate these clusters. This type of exploratory data analysis questions can inform the definition of a new hypothesis for subsequent studies on this dataset and can be facilitated by interacting with the various views provided here. 
For instance, the ring type parallelizes the cluster structure revealed in the self-organizing layout exactly. Various other interesting attributes shed light on how the clusters are formed. We have included a supplementary video that demonstrates some of the interesting attributes that distinguish the clusters revealed by the proposed analysis and visualization technique. 

\begin{figure*}[h!t]
  \begin{center}
  \begin{tabular}{@{\hspace{0pt}}c@{\hspace{50pt}}c@{\hspace{0pt}}}
    \includegraphics[width=0.25\textwidth]{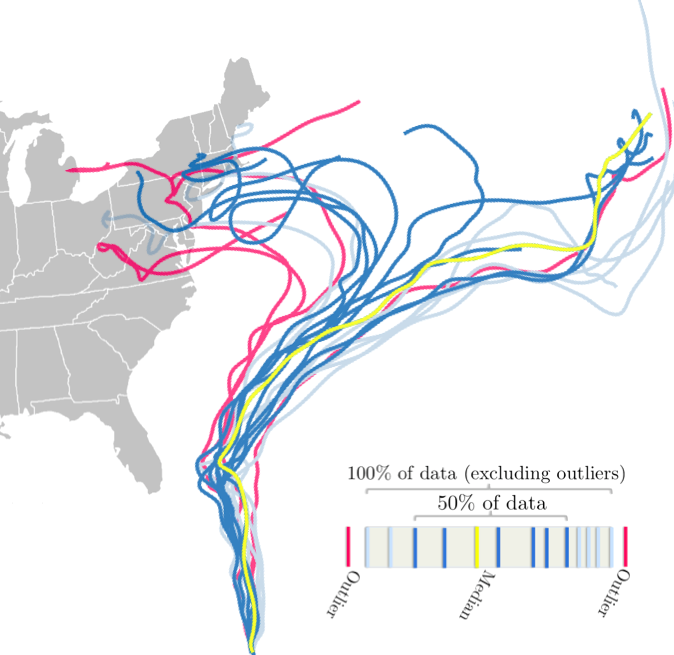} & 
     \includegraphics[width=0.25\textwidth]{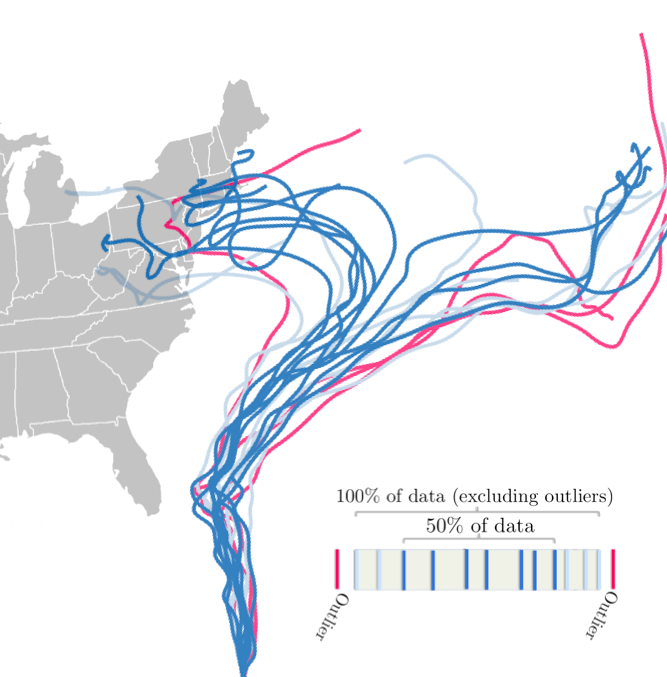}\\  
    \end{tabular}
    \vspace{-7pt}
      \caption{\label{fig:sandy} Hurricane Sandy NCEP GFS ensemble. Left: data depth coloring based on the track information as proposed in~\cite{Mirzargar14}. Right: Data depth coloring based on similarity analysis using all the available attributes.}  
      \end{center}
  \vspace{-15pt}      
\end{figure*}
\begin{figure}[h]
  \begin{center}
    \includegraphics[width=0.4\textwidth]{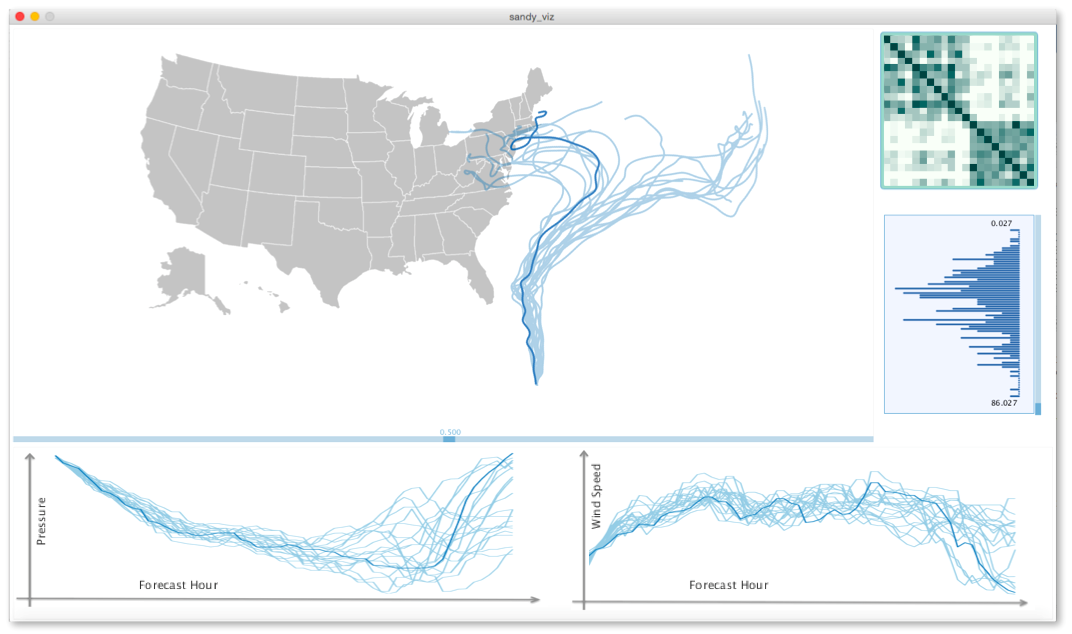} \\
    \vspace{-5pt}
      \caption{\label{fig:sandy2} Brushing and linking can provide information about the hurricane track as well as other available attributes.}
      \end{center}
      \vspace{-15pt}
\end{figure}
In order to provide a comparison to benchmark methods, we used hamming distance or Euclidean distance (depending on the datatype) along with multidimensional scaling. As discussed in Section~\ref{sec:bg1}, MDS is one of the prominent techniques for locating structures and trends in a high-dimensional dataset given a measure of (dis)similarity between datapoints in a dataset~\cite{Williams04}. Hamming distance and Euclidean distance are simple and widely used similarity metrics that have been utilized before to find cluster information using visualization~\cite{Perer08}. Therefore, we have compared our results with MDS based on these metrics. 
Figure~\ref{fig:mushroom_mds} (a) demonstrates the heatmap visualization of the pairwise similarities using Euclidean distance and Figure~\ref{fig:mushroom_mds} (b) demonstrates the MDS projection of this dataset into two dimensions using the similarity matrix in Figure~\ref{fig:mushroom_mds} (a). 
Figure~\ref{fig:mushroom_mds} (c-d) also shows the heatmap visualization and MDS projection after retaining only the similarity values above 0.6 (i..e, thresholded distance as done in~\cite{Perer08}). Note that the clustering pattern in the heatmap visualization and the MDS projection looks very different in this case. This behavior demonstrates the sensitivity of 
MDS projection with respect to the threshold value used for traditional measures of similarity, whereas the proposed technique provides robust behavior with respect to the threshold $\tau$ as demonstrated in the previous example. Changing $\tau$ will reveal only the structures present in the data progressively, whereas thresholding in this case can significantly affect the clustering results in the reduced dimensions. 

Our second example demonstrates the utility of the proposed technique for datasets including multivariate curves in hurricane prediction applications. The visualization of ensembles of hurricane tracks falls in the visualization of uncertainty literature. Various competitive visualization schemes have been proposed for this problem~\cite{Ferstl16, Mirzargar14}, but ensembles of hurricane paths can still be considered as heterogeneous and multidimensional datasets. The hurricane simulation models oftentimes not only produce the tracks (i.e., multivariate curves) but also provide various other attributes such as wind speed, wind radius, and pressure along the predicted tracks. 
The state-of-the-art hurricane analysis and visualization techniques are not designed to take full advantage of all this information, but the proposed technique can handle an ensemble of hurricane tracks with their additional attributes. 

The dataset for this example is from the predictions of Hurricane Sandy. Hurricane Sandy is of particular interest to the meteorology community~\cite{Bassill14} because of its unusual and sudden left turn that had a huge impact on the East Coast of the US. The main problem with the prediction of Hurricane Sandy's path was the significant track bifurcation among several well-trusted forecast models~\cite{Bassill14}. 
We retrieved the publicly available data of NCEP GFS ensemble produced as part of the Tropical Cyclone Guidance Project at the National Center for Atmospheric Research (NCAR)~\cite{Vigh16}, which includes 21 hurricane tracks. Each track includes 60 time points with information about the latitude and longitude of track, as well as wind speed and pressure along the track. Therefore, each time point along the track includes four attributes. 
In this example, the dataset is geospatial. Hence, we used the depth coloring proposed in~\cite{Mirzargar14} for this example and Tukey's rule~\cite{Tukey99} to find the potential outliers. 
Figure~\ref{fig:sandy} (left) demonstrates the visualization of this ensemble using data depth analysis but with the hurricane track information as suggested in~\cite{Mirzargar14}. It is interesting to note that the global median or most central example in this case is consistent with the early impression of Hurricane Sandy. 
That is, it will not make a landfall in the US and many of the tracks on the left have been either captured as an outlier or have light colors (i.e., not representative or central to the ensemble). Figure~\ref{fig:sandy} (right) demonstrates the visualization of the ensemble after performing the similarity analysis using all the attributes and restricting the band size. Note that in this visualization, the multimodal nature of the ensemble is better represented. 
In addition, many of the ensemble members that make landfall in the US now have darker colors (i.e., they are more representative). Similar to the previous example, other attributes of this ensemble can be analyzed by brushing and linking as demonstrated in Figure~\ref{fig:sandy2}. 

Note that this dataset is a 24-dimensional dataset with only 21 datapoints (four attributes at 60 time points for each track). The small size of this ensemble compared to its dimensionality will hamper the utility of analysis based on conventional dimensionality reduction techniques.
\section{Conclusion and Future Work}
\label{sec:conc}

In this paper, we introduced a heterogeneous or mixed-type data visualization framework based on a novel concept of similarity that has a close connection to concepts from nonparametric statistical analysis. 
Compared to other techniques that oftentimes can handle only specific datatypes, the proposed definition is based on the quite generic concept of band inclusion, which can handle different datatypes, including multivariate points, functional values, categorical attributes, or any combinations. We showed that the proposed similarity measure can be used in a spectral clustering context to detect potential local features such as the presence of multiple modes in a dataset.  In addition, unlike current data analysis and dimensionality reduction techniques, the parameter involved in the proposed technique is not a weakness but a strength, as it can show the features of a dataset progressively. 
 
We also proposed a visual encoding technique that provides different levels of abstractions for the analysis and exploration of mixed-type or heterogeneous datasets.  
The proposed visual encoding follows Shneiderman's information-seeking mantra~\cite{shneiderman1996eyes} where an overview of a dataset is provided using a heatmap and a similarity-induced layout. The user can explore the finer structure by brushing and linking to more detailed views. We demonstrated the utility of the proposed technique using multiple examples with various datatypes. 
 
Some of the shortcomings of the proposed technique suggest future lines of research. Starting from the data analysis task, the current technique cannot handle missing attributes/values. However, in a more realistic setting, it is not uncommon to have datasets for which some of the attribute values are missing. The definition of a band inclusion concept that is robust to missing data is still missing and could be an interesting future direction. 
Another shortcoming of the current technique is its scalability. The size of the band inclusion signature can grow rapidly as the size of the dataset grows. In such situations, one can use only subsets of random bands to construct the signatures. This approximation can provide reliable information considering the close connection of the band inclusion signature to the concept of data depth analysis. One can also deploy a lossless compression algorithm to encode the binary signatures. Another aspect of the the scalability issue relates to the similarity-induced layout visualization when force-directed layout is used: the overlay of the nodes on top of each other and the huge amount of edge crossing. These shortcomings can be alleviated by replacing the force-directed layout with other visualization schemes that are more scalable, such as aggregation and bundling of the clusters~\cite{Tukey99, Holten09}.



\bibliographystyle{abbrv-doi}

\bibliography{refs}
\end{document}